# Effect of Low Oxygen Annealing on Photoelectrochemical Water Splitting Properties of α-Fe$_2$O$_3$

Yoichi Makimizu,[a,b] Nhat Truong Nguyen,[a]† Hyo-Jin Ahn,[a,c]‡ JeongEun Yoo,[a] Imgon Hwang,[a] Stepan Kment[c] and Patrik Schmuki*[a,c,d]

[a] *Department of Materials Science and Engineering, University of Erlangen-Nuremberg, Martensstrasse 7, D-91058 Erlangen, Germany*

[b] *Steel Research Laboratory, JFE Steel Corporation, 1 Kokan-cho, Fukuyama, Hiroshima 721-8510, Japan*

[c] *RCPTM, Faculty of Science, Palacky University, 17. listopadu 12, 771 46, Olomouc, Czechia*

[d] *Chemistry Department, Faculty of Sciences, King Abdulaziz University, 80203 Jeddah, Saudi Arabia Kingdom*

† Current address: *Department of Chemistry, University of Toronto, 80 St. George Street, Toronto, Ontario M5S 3H6, Canada*

‡ Current address: *German Engineering Research and Development Center LSTME Busan, Affiliate Institute to FA Universität 7 Erlangen, 1276, Jisa-Dong, Gangseo-Gu, Busan 46742, Republic of Korea*

*Corresponding author. Email: schmuki@ww.uni-erlangen.de

Link to the published article:

https://pubs.rsc.org/no/content/articlehtml/2020/ta/c9ta10358a

Keywords: α-Fe$_2$O$_3$; water splitting; oxygen vacancy; anodization; intensity modulated photocurrent spectroscopy

**Abstract**

Photoelectrochemical (PEC) water splitting is a promising method for conversing solar energy into chemical energy stored in the form of hydrogen. Nanostructured hematite (α-Fe$_2$O$_3$) is one of the most attractive materials for highly efficient charge carrier generation and collection due to its large specific surface area and shortening minority carrier diffusion length required to reach the surface. In the present work, PEC water splitting performance of α-Fe$_2$O$_3$ prepared by anodization of thin iron layers on an FTO glass and subsequent annealing in low O$_2$-Ar ambient with only 0.03% O$_2$ was investigated. The key finding is that annealing the anodic nanostructures with low oxygen concentration provides a strongly enhanced PEC performance compared with classic air annealing. The photocurrent of the former at 1.5 V vs. RHE results in 1.1 mA/cm$^2$, being 11 times higher than that of the latter. The enhancement of the PEC performance for α-Fe$_2$O$_3$ annealed in low oxygen atmosphere can be attributed to controlled morphology, Sn doping, and introduction of oxygen vacancies, which contribute to the enhancement of the hole flux from the photogenerated site to the reactive surface and additionally lead to an enhanced hole transfer at the interface between the α-Fe$_2$O$_3$ and the electrolyte. From the obtained results, it is evident that low oxygen annealing is a surprisingly effective method of defect engineering and optimizing α-Fe$_2$O$_3$ electrodes for a maximized PEC water splitting performance.

## 1. Introduction

Hydrogen with zero CO$_2$ emission ranks among the most promising, clean, and renewable energy vectors that can replace fossil fuels, which still represent the main energy source but have a detrimental impact on the environment. Photoelectrochemical (PEC) water splitting is a promising strategy for conversing solar energy into chemical energy stored in the form of hydrogen.[1] In 1972, Fujishima and Honda first demonstrated the electrochemical photolysis of water into hydrogen and oxygen by using $TiO_2$ as a semiconductor photoanode.[2] Since that seminal work a large range of metal oxide semiconductors have been investigated for their PEC water splitting application.[3-6] Among them, hematite iron oxide (α-$Fe_2O_3$) is one of the most attractive materials due to many favorable properties such as a suitable band gap (1.9-2.2 eV) for absorbing a significant portion of the visible light, high chemical stability in many electrolytes, stability against the photocorrosion, abundance, and low cost.[7,8] Nevertheless, the PEC performance of α-$Fe_2O_3$ is still limited by several key factors such as short excited state lifetime (<10 ps),[9] short diffusion length (2-4 nm) of minority charge carriers (holes),[10] low carrier mobility ($10^{-2}$ to $10^{-1}$ $cm^2/Vs$),[11,12] and a poor electrical conductivity ($10^{-14}$ S/cm),[13] hinder its practical application. These detrimental optical and electronic properties lead to a poor collection of photogenerated holes and their fast recombination with electrons in the hematite bulk. In order to enhance the PEC performance of α-$Fe_2O_3$ photoanodes, a wide range of approaches have been explored, including careful nanostructuring,[14] elemental doping (*e.g.*, by Sn, Ti, Si),[15-17] and the decoration of the hematite surface by various oxygen evolution reaction (OER) co-catalysts (*e.g.*, $IrO_2$, Co-Pi (Co-phosphate), FeOOH).[18-20] Sivula *et al.*[15] reported that nanocrystalline α-$Fe_2O_3$ photoanodes deposited on an FTO glass (transparent fluorine doped tin oxide layer coated on a glass substrate) that are annealed at 700 °C or below exhibit no photocurrent for PEC water splitting. However, upon the annealing at 800 °C, the photoactivity of α-$Fe_2O_3$ was drastically improved as a result of the thermally driven diffusion of Sn ions from the FTO substrate into the hematite lattice acting here as the donor dopant. Similarly, the introduction of oxygen vacancies has also been reported to enhance the photoresponse of α-Fe$_2$O$_3$ *via* a self-doping mechanism.[21] Furthermore, nanostructuring of α-Fe$_2$O$_3$ provides a highly improved efficiency in the charge carrier generation and collection due to the enhancement of the specific surface area and drastic shortening of the minority carrier diffusion length.[14] The α-Fe$_2$O$_3$ nanostructures have been synthesized using a variety of techniques including sol-gel processing,[22] electrodeposition,[23] spray pyrolysis,[24] hydrothermal synthesis,[25,26] magnetron sputtering,[27,28] and electrochemical anodization.[29,30] Among them, anodization is considered a viable technique for the fabrication of nanostructured α-Fe$_2$O$_3$ from the viewpoint of large-scale production and cost reduction.[30] The anodically fabricated oxide films are typically amorphous and need to be thermally treated in order to become converted into the crystalline form (hematite α-Fe$_2$O$_3$). However, if the nanostructured layers are formed on metallic iron substrates, the thermal annealing leads to the formation of gradient oxide layers consisting of wustite (FeO), magnetite (Fe$_3$O$_4$), and α-Fe$_2$O$_3$ crystalline phases.[31] The presence of these phases cannot be avoided because of the complex additional thermal oxidation of underlying metal substrate during the annealing. The FeO and Fe$_3$O$_4$ phases are not suitable for a photolysis of water as they behave either metal-like or as a narrow band gap (<1 eV) semiconductor,[21,29] therefore, plain α-Fe$_2$O$_3$ is the desired phase for PEC water splitting applications. A complete anodization of a magnetron sputtered metallic iron film on the FTO substrate prior to the thermal annealing is an efficient strategy to avoid the creation of the unwanted iron oxide phases.[32,33] Furthermore, the preparation of photoanodes on a transparent substrate enables back side illumination, *i.e.*, the irradiation of the sample from the direction of the substrate/semiconductor interface, which, in many cases, have led to an improved PEC activity.[34,35] It is also well known that annealing conditions such as temperature and atmosphere affect the PEC performance of α-Fe$_2$O$_3$ layers.[15,30,32,36-38] Ling *et al.*[36] reported that photoresponse of hydrothermally grown α-Fe$_2$O$_3$ nanowires was activated by annealing in an oxygen deficient atmosphere achieved in an evacuated furnace refilled by pure $N_2$. It was shown that the generation of oxygen vacancies considerably increased the electrical conductivity. On the other hand, it was also described that α-$Fe_2O_3$, under these reducing conditions, can be easily converted into undesired $Fe_3O_4$. In the present work, we create a fully converted α-$Fe_2O_3$ nanostructure on FTO substrate by the aforementioned approach. Key is the use of a defined low oxygen annealing environment to obtain a highly active nanostructured α-$Fe_2O_3$ photoanode. First, high-quality iron films with carefully controlled thicknesses were deposited on an FTO substrate by a modified magnetron sputtering technique. Subsequently, the films were fully anodized to amorphous nanostructured iron oxide. Finally, we investigated the effect of annealing of these nanostructures on their PEC performance in a defined gas environment containing a very low concentration of oxygen (Ar + 0.03% $O_2$ atmosphere). We demonstrate that the controlled annealing in low oxygen ambient can drastically improve the PEC performance of α-$Fe_2O_3$ photoelectrodes (more than tenfold compared with classic air annealing).

## 2. Experimental section

*2.1 Preparation of α-$Fe_2O_3$ layers*

The α-$Fe_2O_3$ photoanodes were prepared by anodization of a thin Fe layer deposited on FTO glass supports, followed by ambient controlled annealing. Initially, thin Fe layers were prepared on an FTO glass (TOC22-15, Solaronix) by a high-energy magnetron sputtering method known as high power impulse magnetron sputtering (HiPIMS). Before the deposition, the FTO substrates were carefully cleaned by subsequent sonication baths of isopropyl alcohol, ethanol, and water. The HiPIMS depositions were carried out in an ultra-high vacuum (UHV) reactor continuously pumped down by turbo-molecular pump providing a base pressure of $10^{-5}$ Pa. The metallic target of pure iron (99.995%, Plasmaterials) with an outer diameter 50 mm and an argon atmosphere were used as the iron source and a working gas, respectively. The working gas was fed to the reactor with the flow rate of 30 sccm (standard cubic centimeters per minute). The FTO substrates were not externally heated during the depositions, so the temperature of the FTO rose up to ~ 80 °C owing to the energy bombardment. The operating pressure was 1 Pa. The pulsing frequency of the DC HiPIMS discharge was 100 Hz with the duty cycle of 1 %. The sputtering time was adjusted so that the thickness of the deposited Fe layers on FTO was 300 nm.

In the next step, the Fe films were transformed into the amorphous iron oxide nanostructures by electrochemical anodization procedure. The anodization was performed in a solution of ethylene glycol (EG, ≥99.5%, Carl Roth) containing 0.2 M $NH_4F$ (≥98%, Carl Roth) and 3 vol% $H_2O$ at 50 V and at room temperature. A two-electrode system, in which the Fe layer on the FTO and a Pt sheet served as the working and counter electrodes, respectively, was used. After the anodization the samples were rinsed with water and dried in a nitrogen stream. The anodized layers were then annealed at 600 °C for 40 min and 750 °C for 20 min in air and in 0.03% $O_2$-Ar ambient using a tube furnace (Linn High Therm, FRH-40/250/1500). The 0.03% $O_2$-Ar ambient was provided by a continuous flow through the furnace using a commercial gas cylinder (VARIGON® S, Linde) prior to and during the annealing process. The samples were labeled based on the annealing temperature-atmosphere (*i.e.*, 600-Air, 750-LO (Low Oxygen)).

*2.2 Layer characterization*

The morphology of layers was investigated using a scanning electron microscope (SEM, Hitachi, S-4800). X-ray diffraction (XRD) was performed with an X'pert Philips MPD (equipped with a Panalytical X'celerator detector) with a graphite monochromatic $Cu_{K\alpha}$ radiation (λ =

1.54056 Å). The chemical composition and the oxidation state were characterized by X-ray photoelectron spectroscopy (XPS, PHI 5600), and the peak positions were calibrated on the C 1s peak at 284.8 eV. Solid-state conductivity measurements were carried out in an adapted scanning electron microscope (SEM, TESCAN LYRA3 XMU) by 2-point measurements using a semiconductor characterization system (Keithley 4200-SCS). Tungsten tips were used as electrical contacts between the α-$Fe_2O_3$ layer surface and the FTO substrate. Resistivity values were then obtained from the *I-V* curves by ramping the potential from -2 V to 2 V at a sweep rate of 5 mV/s.

*2.3 Photoelectrochemical measurements*

The PEC performance of the α-$Fe_2O_3$ photoanodes was measured in a three-electrode PEC cell, where a Pt counter electrode and a Ag/AgCl (3M KCl) reference electrode in a 1.0 M KOH electrolyte were used. The photocurrent-potential (*J-V*) properties were studied by scanning the potential from -0.5 to 0.7 V at a scan rate of 2 mV/s under periodic illumination of AM 1.5 G (100 mW/$cm^2$) light. The potentials versus Ag/AgCl (3 M KCl) were converted to the reversible hydrogen electrode (RHE) according to the Nernst equation.

$$E_{\text{RHE}} = E_{\text{Ag/AgCl}} + E^0_{\text{Ag/AgCl}} + 0.059\ \text{pH} \qquad (1)$$

where $E_{\text{RHE}}$ is the converted potential versus RHE, $E_{\text{Ag/AgCl}}$ is the experimentally measured potential, and $E^0_{\text{Ag/AgCl}} = 0.209$ V at 25 °C for a Ag/AgCl electrode in 3 M KCl. The incident photon to current efficiency (IPCE) was acquired in the range from 300 to 700 nm with 5 nm steps at an applied potential of 1.5 V vs. RHE in 1.0 M KOH. Electrochemical impedance spectroscopy (EIS), Mott-Schottky measurements, and intensity modulated photocurrent spectroscopy (IMPS) measurements were carried out using a Zahner IM6 (Zahner Elektrik) with a tunable light source TLS03. The EIS measurements were carried out in the frequency range from 100 kHz to 0.1 Hz at 1.5 V vs. RHE with a perturbation amplitude of 10 mV and a 452 nm light source with the intensity of 10 mW/cm$^2$. The Mott-Schottky measurements were conducted at a frequency of 10 kHz under dark conditions. The donor density ($N_d$) was calculated by the following equation (2).

$$N_\mathrm{d} = (2/e_0\varepsilon\varepsilon_0)[\mathrm{d}(1/C^2)/\mathrm{d}V]^{-1} \qquad (2)$$

where $e_0$ is the electron charge (1.60 × 10$^{-19}$ C), $\varepsilon$ is the dielectric constant of α-Fe$_2$O$_3$ (80),[17] $\varepsilon_0$ is the permittivity vacuum (8.85 × 10$^{-12}$ Fm$^{-1}$), and $C$ is the capacitance derived from the electrochemical impedance at each potential ($V$). The IMPS responses were recorded in the range of 1.0–1.7 V vs. RHE with 0.1 V steps under 452 nm light illumination with the intensity of 10 mW/cm$^2$. The light intensity was modulated by 10% between 1 kHz and 0.1 Hz.

## 3. Results and discussion

In order to produce nanostructured phase-pure α-Fe$_2$O$_3$ photoanodes without forming Fe$_3$O$_4$ and/or FeO phase impurities, a thin metallic Fe layer with a thickness of 300 nm was deposited on an FTO glass by magnetron sputtering. Next, the Fe layer was anodized at room temperature in an EG electrolyte containing 0.2 mol/L NH$_4$F and 3 vol% H$_2$O using the voltage of 50 V. The end point of the anodization was detected by an increase in the anodic current recorded on-line during the anodization process (Fig. S1).[32] The anodization procedure completely transformed the non-transparent iron layer to fully transparent iron oxide nanostructure as it is evident from Fig. 1(a-1). Fig. 1(b) shows the surface SEM image of the layer after anodization, and the morphology exhibits a nanoporous structure. Subsequently, the anodized layers were annealed at 600 and 750 °C in air and 0.03% O$_2$-Ar ambient. All the annealed layers had a thickness of 450-520 nm. The surface appearance and SEM images of these layers are shown in Fig. 1(a-2, c-f). The layers were transparent even after the annealing. However, the color of the layers changed from yellow to red after the anodization and subsequent annealing, as it is apparent for the α-Fe$_2$O$_3$ photoanode annealed at 600 °C in low oxygen ambient (600-LO) (Fig. 1(a-2)); other samples also exhibited a similar red color, which indicated a transformation of the amorphous structure to the α-Fe$_2$O$_3$ phase.[39] The morphologies of these layers changed depending on the annealing conditions. Larger particles were formed in the layers annealed at 750 °C (750-LO, 750-Air) compared with those annealed at 600 °C (600-LO, 600-Air) under the same ambient conditions. Similar coarsening of the particles at higher temperatures has also been reported in literature.[15] Interestingly and importantly, the coarsening of the particles was inhibited in low oxygen annealing (600-LO, 750-LO) compared with air annealing (600-Air, 750-Air). In the case of the 600-LO sample, the nanoporous structure obtained during the anodization was fully maintained. The SEM images verified that the lower temperature and low oxygen concentration annealing were able to retain the desired nanostructured morphology.

The crystal structure of the layers after the annealing process was determined by XRD, and the resulting XRD patterns are shown in Fig. 2. Clearly, only the peaks corresponding to α-Fe$_2$O$_3$ and SnO$_2$ attributed to the FTO substrate were identified. In other words, a layer that fully consists of only α-Fe$_2$O$_3$ can be fabricated by completely anodizing an iron thin layer on the FTO glass followed by annealing.

The layers were further analyzed by XPS. Fig. 3 shows the survey spectra as well as the high-resolution Sn 3d, O 1s, and Fe 2p spectra of the α-Fe$_2$O$_3$ after annealing under various conditions. The survey spectra revealed that the α-Fe$_2$O$_3$ contained not only Fe and O but also Sn dopants due to the diffusion from the FTO substrate during the annealing process (Fig. 3(a)).[15,33] The high resolution XPS spectra of the Sn 3d indicate the increase in the Sn concentration after annealing at a higher temperature and in low oxygen ambient (Fig. 3(b)). The atomic concentrations of the Sn quantified from XPS spectra are shown in Table 1. Remarkably, the α-Fe$_2$O$_3$ annealed in low oxygen environment shows a higher Sn concentration compared with air annealing. This can be attributed to the acceleration of Sn diffusion from the FTO substrate into the α-Fe$_2$O$_3$ by grain boundary diffusion since the α-Fe$_2$O$_3$ annealed in the low oxygen ambient provides a fine crystalline nanostructure, as shown in Fig. 1. The spectra of O 1s show that the binding energy centered at around 529.7 eV for 600-Air is shifted to a slightly higher binding energy compared with the α-Fe$_2$O$_3$ samples having a higher Sn concentration (Fig.3(c)). This suggests that the Sn doping would exert influence on the Fe-O lattice and increase the binding energy.[40] In the Fe 2p spectra, Fe 2p$_{3/2}$ peaks around 711 eV, Fe 2p$_{1/2}$ peaks around 724 eV, and satellite peaks of Fe$^{3+}$ around 719 eV can be clearly identified (Fig. 3(d)). These values have been reported as typical binding energies for Fe$_2$O$_3$.[36,39,41-43] In addition to these peaks, a variation of peak intensity at 716 eV, which corresponds to a Fe$^{2+}$ satellite peak,[36,39,41-43] can be observed, depending on the annealing conditions. Peak deconvolution was performed on each Fe 2p spectrum (Fig. S2),[44,45] and the Fe$^{2+}$/Fe$^{3+}$ ratio was obtained from the area ratio of each satellite peak, as shown in Table 1. The ratio of Fe$^{2+}$ increased for low oxygen and higher temperature annealing. These results suggest that oxygen vacancies that act as shallow donors are created in the α-Fe$_2$O$_3$.[36] Since the Sn dopants as well as the oxygen vacancies can act as an electron donor, they can improve the electrical conductivity of α-Fe$_2$O$_3$ *via* a polaron hopping mechanism.[46,47]

In order to confirm the conductance of the α-Fe$_2$O$_3$ layers, solid state conductivity measurements were carried out in an adapted SEM by 2-point measurements using a semiconductor characterization system. The measured *I-V* curves are shown in Fig. S3 and the corresponding electric resistances are shown in Table 1. The α-Fe$_2$O$_3$ annealed in the low oxygen ambient (750-LO, 600-LO) shows a higher conductivity compared with samples annealed in air (750-Air, 600-Air). This can be ascribed to the increased donor density by Sn doping and the introduction of oxygen vacancies. Therefore, annealing in low oxygen ambient is an effective method for enhancing electrical conductivity of α-$Fe_2O_3$. However, it should be noted that 600-LO shows the best conductivity despite the fact that 750-LO has the highest Sn concentration and $Fe^{2+}$ ratio. Yang *et al.*[48] reported on the conductance of α-$Fe_2O_3$ doped with various concentrations of Sn and revealed that the electrical conductance of α-$Fe_2O_3$ can be improved by increasing the Sn concentration only up to a certain critical concentration level (in fact, corresponding to the solubility limit of Sn in $Fe_2O_3$); above this level the overall conductivity decreases significantly. Similarly, in the case of 750-LO in our experiment, Sn seemed to act as a scattering center that reduced the conductivity of α-$Fe_2O_3$ once the solubility of Sn was reached. Therefore, exceeding the optimal annealing temperature (with excessively high Sn concentrations) leads to the conductivity deterioration of α-$Fe_2O_3$.

The PEC water splitting performance of the fabricated α-$Fe_2O_3$ layers was measured in a 1.0 M KOH electrolyte. The corresponding photocurrent-potential (*J-V*) curves recorded under the chopped light illumination (AM 1.5G, 100 mW/$cm^2$) are shown in Fig. 4(a). The 600-LO sample that shows preserved nanostructuring and has the highest conductivity also exhibits the best PEC water splitting performance. The photocurrent value reaches 1.1 mA/$cm^2$ at 1.5 V vs. RHE, which is 1.8 times and 11 times higher than that of 750-Air (0.6 mA/$cm^2$) and 600-Air (0.1 mA/$cm^2$) samples, respectively. Further, the quantum yields of the reactions represented as the IPCE values were measured at 1.5 V vs. RHE as a function of the incident light wavelength; the results are available in Fig. 4(b). The IPCE values for all α-$Fe_2O_3$ layers gradually drop to zero at wavelengths above 610 nm, in accordance with the bandgap of α-$Fe_2O_3$.[7] Importantly, the 600-LO shows significantly enhanced IPCE magnitude compared with all the other α-Fe2O3 layers over the entire wavelengths that are in accordance with the J-V curves. The highest efficiency is 18.8% at the wavelength of 330 nm. Literature[15,49,50] mentioned that the PEC performance of α-Fe$_2$O$_3$ photoanodes on an FTO glass was greatly improved by their annealing at above 750 °C or at a lower temperature for a long time (*e.g.*, 600 °C for 8 h). The enhancement of the PEC performance was attributed to the increased electrical conductivity as a result of the incorporation of Sn as the donor doping element coming from the FTO substrate due to the thermal diffusion. This phenomenon occurred in our samples annealed in air - we observed a better performance for the 750-Air sample, which provided a higher Sn concentration and conductivity compared with the 600-Air sample. In contrast, the results for low oxygen annealing (600-LO, 750-LO) exhibited the best PEC performance for the α-Fe$_2$O$_3$ annealed at 600 °C for 40 min. The conversely degraded PEC performance for 750-LO can be attributed to a combination of larger particle size of α-Fe$_2$O$_3$ (Fig. 1) and lower conductivity (Table. 1) compared with 600-LO. This indicates that the enhancement of the photoresponse of the α-Fe$_2$O$_3$ annealed in low oxygen ambient can be activated even at lower temperature of 600 °C and by shorter annealing times. Annealing at a lower temperature and simultaneously for shorter time gains practical significance because it can reduce the damage to the structure and electrical conductivity of the FTO substrate as well as reducing the energy cost of the annealing process.[51]

In order to investigate the donor densities of α-Fe$_2$O$_3$ layers annealed under every conditions, Mott-Schottky measurements were carried out in 1.0 M KOH electrolyte under dark conditions. The Mott-Schottky plots are shown in Fig. 5(a). The slopes of Mott-Schottky plots for all α-Fe$_2$O$_3$ samples are positive, which indicates that they are n-type semiconductors with electrons as majority carriers. The donor densities estimated from the slopes of Mott-Schottky plots are shown in Table 1. Importantly, the 600-LO sample, which exhibited the best PEC water splitting performance, showed the highest donor density of $4.88 \times 10^{+19}$, which was more than one order higher than that of the 600-Air sample ($4.86 \times 10^{+18}$). Since the donor density of α-Fe$_2$O$_3$ can be increased by Sn doping and the introduction of oxygen vacancies,[36,49] the improvement of the donor density for the 600-LO can be attributed to these effects. The donor densities are also consistent with the conductivity measurements (Table 1 and Fig. S3). In other words, low oxygen annealing promotes Sn doping and introduction of oxygen vacancies, and increases the donor density of α-Fe$_2$O$_3$. As a result, the improvement of electrical conductivity of α-Fe$_2$O$_3$ leads to an excellent PEC water splitting performance. However, it should be noted that 750-LO shows a lower donor density than that of 600-LO in spite of exhibiting the highest Sn concentration and Fe$^{2+}$ ratio. Aroutiounian et al.[52] investigated the donor density of Sn doped α-Fe$_2$O$_3$ and suggested that the donor density of α-Fe$_2$O$_3$ drastically increased until 1.0 at% of Sn, and then decreased sharply to 1.5 at%. As described above for the change in the conductivity of Sn doped α-Fe$_2$O$_3$, it has been reported that the donor density and the conductivity of α-Fe$_2$O$_3$ decrease when the Sn concentration exceeds a certain level. Our results agree well with these findings.

Furthermore, EIS measurements of the α-Fe$_2$O$_3$ layers were carried out in a 1.0 M KOH electrolyte at 1.5 V vs. RHE under visible light illumination (wavelength of 452 nm, 10 mW/cm$^2$). The EIS results in the form of Nyquist plots are shown in Fig. 5(b). The experimental data were fitted by using the equivalent circuit model depicted in the inset image of Fig. 5(b), and the fitting results are summarized in Table S1. The equivalent circuit consists of a series resistance, $R_s$, bulk resistance of α-Fe$_2$O$_3$, $R_1$, charge transfer resistance at the α-Fe$_2$O$_3$/electrolyte interface, $R_2$, the space charge capacitance of the bulk α-Fe$_2$O$_3$, $C_1$, and the space charge capacitance at the α-Fe$_2$O$_3$/electrolyte interface, $C_2$.[53,54] Sample 600-LO exhibits the smallest semicircle and smallest values of $R_1$ and $R_2$ compared with all other α-Fe$_2$O$_3$ layers. The smallest value of $R_1$ for 600-LO coincides with the results of the conductivity and the Mott-Schottky measurements, which show the highest conductivity and the highest donor density, respectively, for 600-LO. In other words, the electrons, which are majority carriers of the α-Fe$_2$O$_3$ electrode, can be rapidly transferred within the bulk α-Fe$_2$O$_3$, which has a high conductivity. This, in turn, reduces their recombination with photoexcited holes. Therefore, $R_1$ can be assigned to the recombination resistance inside the bulk α-Fe$_2$O$_3$, and the smallest $R_1$ for 600-LO denotes a reduced recombination and an easy extraction of the holes from the bulk α-Fe$_2$O$_3$ through the electrode surface. On the other hand, a smaller value of $R_2$, which means charge transfer resistance, indicates that the holes extracted into the electrode surface easily transfer from α-Fe$_2$O$_3$ to the electrolyte and contribute to the OER, which is in line with literature for Sn doped α-Fe$_2$O$_3$.[55-57] Therefore, the smallest $R_2$ for 600-LO can be attributed to Sn doping into the α-Fe$_2$O$_3$.

IMPS measurements for 600-LO and 600-Air were carried out in order to understand the effect of low oxygen annealing on the kinetics of electron-hole recombination and hole transfer in greater detail. The variations of IMPS responses as a function of applied potential were measured in a 1.0 M KOH electrolyte under illumination (wavelength of 452 nm, 10 mW/cm$^2$). The theoretical background to IMPS measurements was described in-depth by Peter *et al.*[58-61] Fig. 6(a) shows a typical IMPS response for 600-LO at 1.1 V vs. RHE. Two semicircles can be identified in the complex plane plots. The lower semicircle corresponds to the high frequency part that arises from RC attenuation by series resistance of the FTO glass and the space charge capacitance of the α-Fe$_2$O$_3$ layer, whereas the upper semicircle corresponding to the low frequency part indicates the competition between the recombination and hole transfer. Therefore, the rate constant of recombination, $k_{rec}$, and the rate constant of hole transfer, $k_{trans}$, can be evaluated by analyzing the upper semicircle. Typically, when the illumination is irradiated to α-Fe$_2$O$_3$ electrode, instantaneous photocurrent is observed, which corresponds to a hole current. The instantaneous hole current is not associated with charge transfer across the interface between electrode/electrolyte. Under the continued illumination, the photocurrent exponentially decays due to the hole build-up and the recombination of the holes and electrons until it reaches steady-state photocurrent (Fig. S4). In the IMPS response shown in Fig. 6(a), the maximum real photocurrent at high frequency, $j_{HF}$, and the minimum real photocurrent at low frequency, $j_{LF}$, correspond to the instantaneous photocurrent when the light is irradiated and the steady-state photocurrent under illumination, respectively. The ratio of these photocurrents represents the transfer efficiency, $\eta_{trans}$, which is the fraction of holes that successfully transferred to the electrode/electrolyte interface and contributed to the OER, as shown in equation (3).

$$\eta_{trans} = \frac{j_{LF}}{j_{HF}} = \frac{k_{trans}}{k_{trans}+k_{rec}} \qquad (3)$$

Furthermore, the radial frequency at the maximum point of the upper semicircle corresponds to the sum of $k_{trans}$ and $k_{rec}$ as expressed in equation (4).

$$\omega_{max} = k_{trans} + k_{rec} \qquad (4)$$

Consequently, $k_{trans}$ and $k_{rec}$ can be derived from equation (3) and (4). The variations of $j_{HF}$ and $j_{LF}$ for 600-LO and 600-Air as a function of applied potential are shown in Fig. 6(b). The variation of $j_{LF}$ shown by open symbols in Fig. 6(b) is consistent with the *J-V* curves displayed in Fig. 4(a) because both of these photocurrents represent the situation under steady-state photocurrents. Importantly, 600-LO exhibits a greatly higher $j_{HF}$ over the entire potential range compared with 600-Air conditions. Considering that the $j_{HF}$ correspond to the hole flux without recombination, this trend thus indicates that the photogeneration and transport of holes are significantly improved for 600-LO. Since nanostructuring reduces the diffusion length required for holes to reach the electrode surface, in addition to improving the light harvesting, it is plausible that the nanostructure morphology of 600-LO (see Fig. 1(c)) contributes to the increase of $j_{HF}$. Fig. 6(c) shows the variation of $\eta_{trans}$ for 600-LO and 600-Air derived by equation (3). Also in this case the $\eta_{trans}$ for 600-LO exhibits much higher values compared with 600-Air. The increase in the transfer efficiency, $\eta_{trans}$, can be explained in terms of the competition between the interfacial hole transfer and the electron-hole recombination at the surface. In other words, a higher $\eta_{trans}$ means an enhancement of the interfacial hole transfer and/or suppression of the electron-hole recombination at the surface. Therefore, in order to discuss the kinetics of hole transfer and recombination, the variations of both rate constants as a function of potential in the range of 1.0-1.3 V vs. RHE are summarized in Fig. 6(d). It should be noted that the extraction of reliable rate constants becomes difficult for the potential of equal or greater than1.4 V vs. RHE as the upper semicircles for 600-LO become extremely small. The rate constants of surface recombination, $k_{rec}$, depend on the applied potential for both α-$Fe_2O_3$ layers, which is expected to be related to the electron concentration at the surface of α-$Fe_2O_3$.[62] As the rate constants of recombination for 600-LO clearly exhibit smaller values compared with the 600-Air sample, this suggests suppression of electron-hole recombination at the surface. EIS measurements indicated suppression of recombination in the bulk α-$Fe_2O_3$, as described above. The suppression of recombination both on the surface and in the bulk of α-$Fe_2O_3$ can be attributed to the improvement in the conductivity for 600-LO due to Sn doping and the introduction of oxygen vacancies in α-$Fe_2O_3$. The rapid transport and transfer of electrons leads to a reduction in the recombination rate of the photoexcited holes with electrons. The rate constants of hole transfer, $k_{trans}$, for 600-LO exhibit higher values compared with 600-Air. As shown in Table S1, EIS measurements revealed the smallest value of $R_2$ (a charge transfer resistance at the α-$Fe_2O_3$/electrolyte interface) for 600-LO. Both results suggest a significant enhancement of the hole transfer at the interface between the α-Fe$_2$O$_3$ and the electrolyte. Dunn *et al.*[61] also reported an increase in the rate constant of the hole transfer for the Sn doped α-Fe$_2$O$_3$. Thus, the enhancement of the hole transfer for 600-LO can be attributed to Sn doping in the α-Fe$_2$O$_3$, indicating an effective catalysis of the OER on the surface of the α-Fe$_2$O$_3$. It can be concluded from the IMPS measurements that the low oxygen annealing improves the hole flux from the photogenerated site to the reactive surface because of the nanostructuring of α-Fe$_2$O$_3$. In addition, the electron-hole recombination at the surface of α-Fe$_2$O$_3$ is suppressed due to the improvement in the conductivity of α-Fe$_2$O$_3$, and the hole transfer at the interface between the α-Fe$_2$O$_3$ and the electrolyte is enhanced due to the Sn doping in α-Fe$_2$O$_3$.

4. **Conclusions**

In the present study, we investigated the PEC water splitting performance of nanostructured α-Fe$_2$O$_3$ photoanode prepared on an FTO glass and particularly the effect of annealing in a low oxygen content environment, which was 0.03% O$_2$-Ar ambient. The α-Fe$_2$O$_3$ layer produced by anodization of sputtered metallic iron film followed by annealing at 600 °C for 40 min in low oxygen ambient exhibits a strongly enhanced PEC performance compared with conventional air annealing. The photocurrent under illumination AM 1.5 G, 100 mW/cm$^2$ for the α-Fe$_2$O$_3$ annealed in low oxygen ambient resulted 1.1 mA/cm$^2$ at 1.5 V vs. RHE, which was 11 times higher than that of the air annealed reference. From the results of multiple analyses and evaluations, the following mechanistic facts can be deduced: The improvement in the PEC performance for α-Fe$_2$O$_3$ annealed in a low oxygen ambient can be attributed to a coordinated action of nanostructuring, Sn doping, and introduction of suitable oxygen vacancies. Low oxygen annealing avoids coarsening of the α-Fe$_2$O$_3$ nanostructure. This allows to maintain short diffusion length of holes, and supports the hole flux from the photogenerated site to the reactive surface. Sn doping from FTO into α-$Fe_2O_3$ and introduction of oxygen vacancies are enhanced by the low oxygen annealing. These effects contribute to an increase in the donor density and result an improvement of the conductivity, which leads to a low electron-hole recombination rate. Additionally, Sn doping into α-$Fe_2O_3$ enhances also the hole transfer at the interface between the electrode and electrolyte. On the basis of these results, it is evident that low oxygen annealing of α-$Fe_2O_3$ nanostructure is a very effective method for enhancing their PEC water splitting performance. It also demonstrates that low oxygen annealing can activate a strong photoresponse of α-$Fe_2O_3$ on FTO glass even at a significantly lower temperature and for shorter times than previously reported, which is of high practical significance.

**Conflicts of interest**

There are no conflicts to declare.

**Acknowledgement**

The authors would like to acknowledge ERC, DFG, the Erlangen DFG cluster of excellence (EAM) and the Operational Programme Research, Development and Education-European Regional Development Fund, Project No. CZ.02.1.01/0.0/0.0/15_003/0000416 of the Ministry of Education, Youth and Sports of the Czech Republic. Y.M. would like to thank JFE Steel Corporation for the support during his sabbatical.

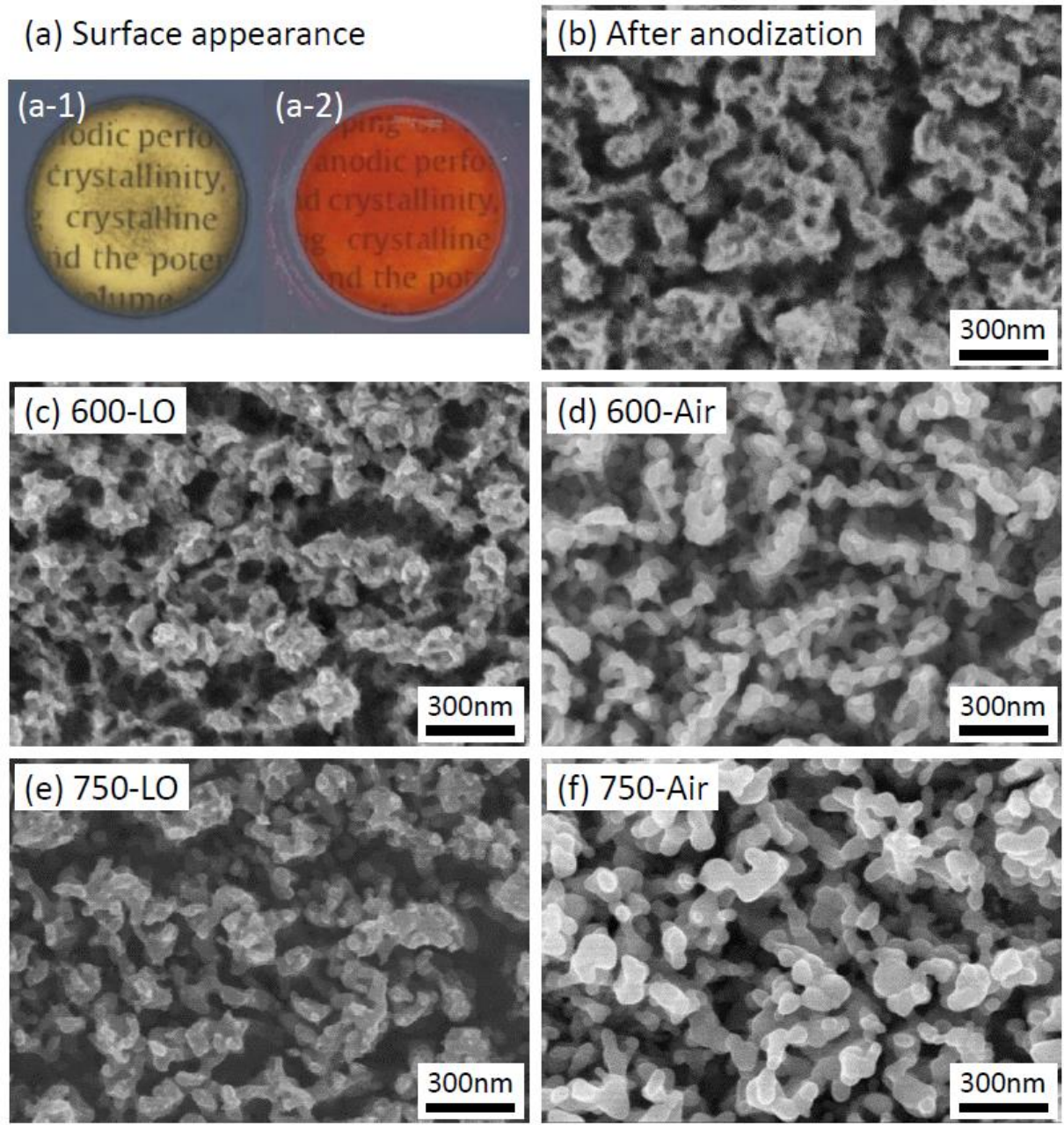

Figure 1. Surface appearance (a-1) after anodization and (a-2) after annealing of 600-LO, and SEM images (b) after anodization and after annealing at (c, d) 600 °C for 40 min and (e, f) 750 °C for 20 min in (c, e) 0.03% $O_2$-Ar and (d, f) air ambient.

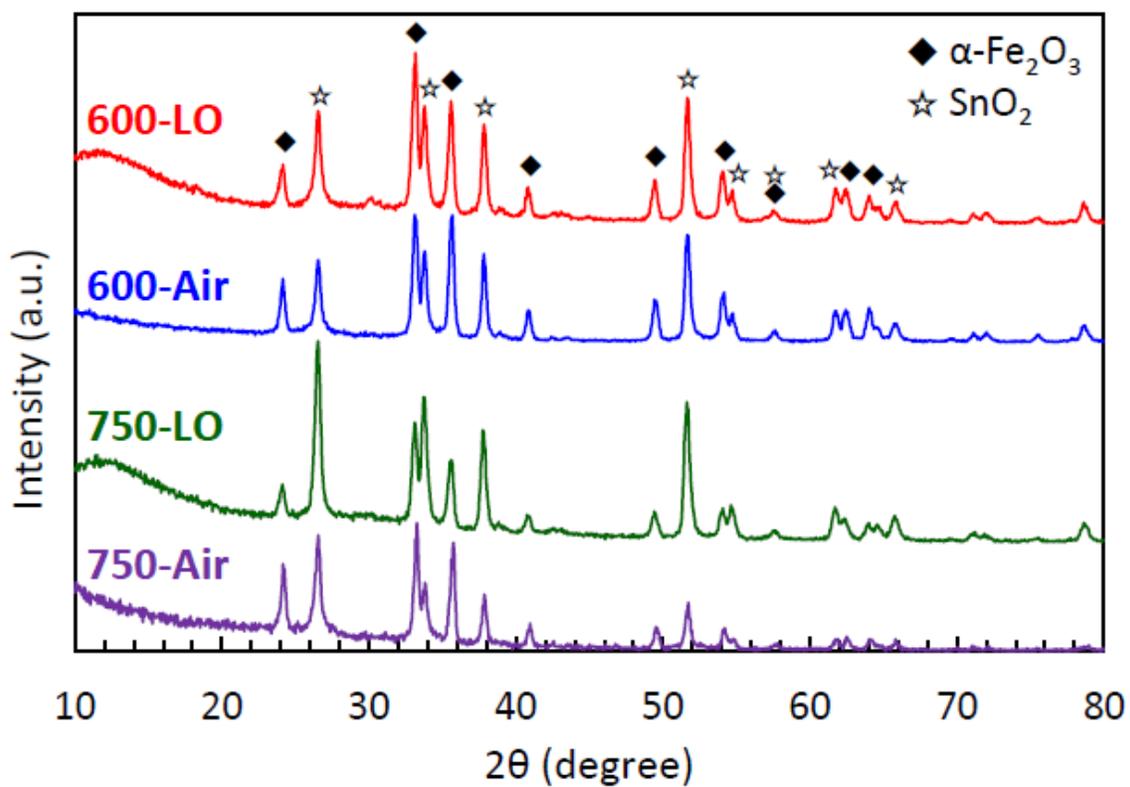

Figure 2. XRD patterns of α-Fe$_2$O$_3$ layers annealed under various conditions.

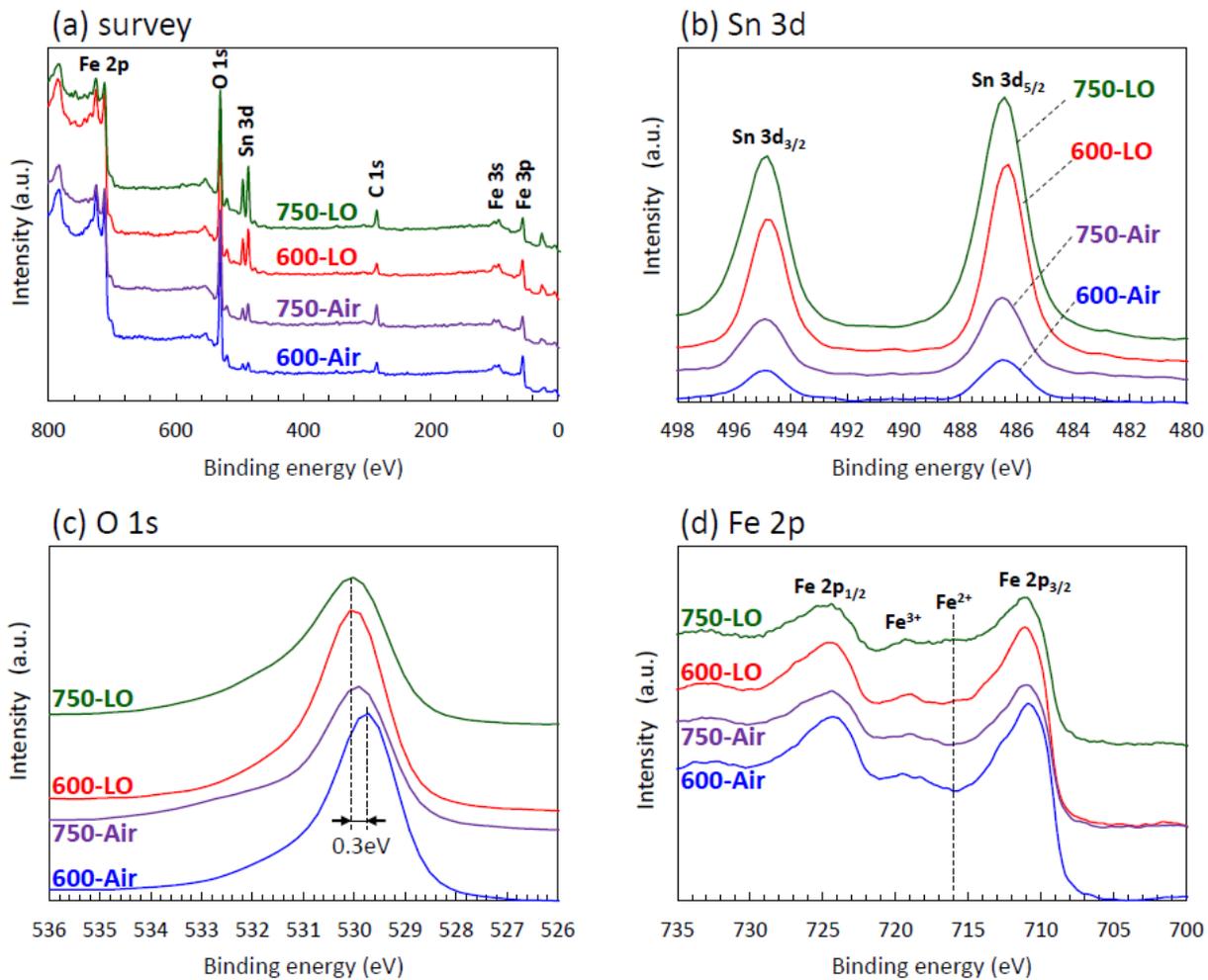

Figure 3. XPS spectra of (a) survey, (b) Sn 3d, (c) O 1s, and (d) Fe 2p for α-Fe$_2$O$_3$ layers annealed under various conditions.

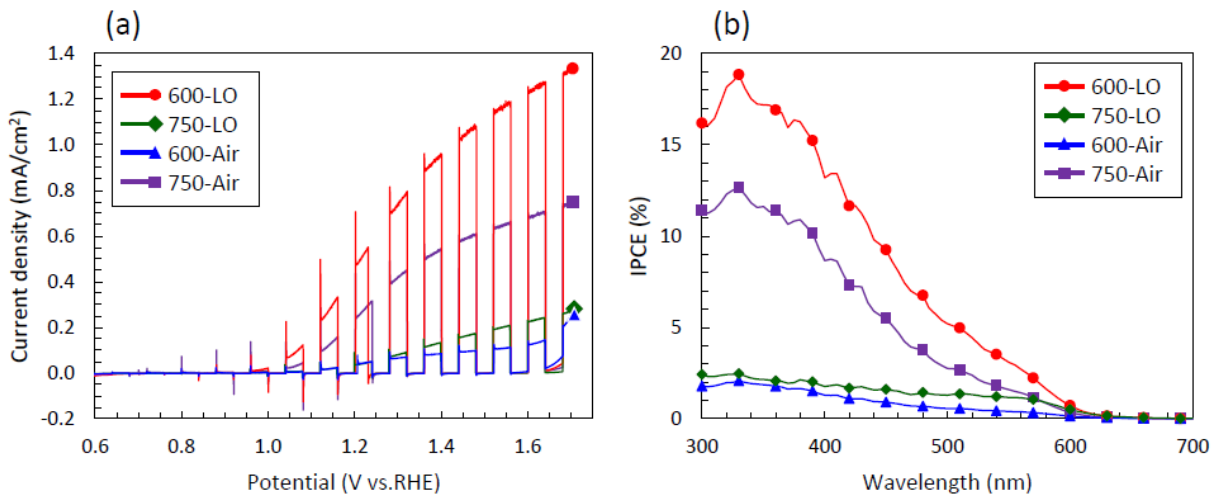

Figure 4. (a) Photocurrent-potential (*J-V*) curves with chopped light illumination (AM 1.5G, 100 mW/cm$^2$) and (b) IPCE spectra measured at 1.5 V vs. RHE for α-Fe$_2$O$_3$ layers annealed under various conditions. All samples were measured in 1.0 M KOH electrolyte.

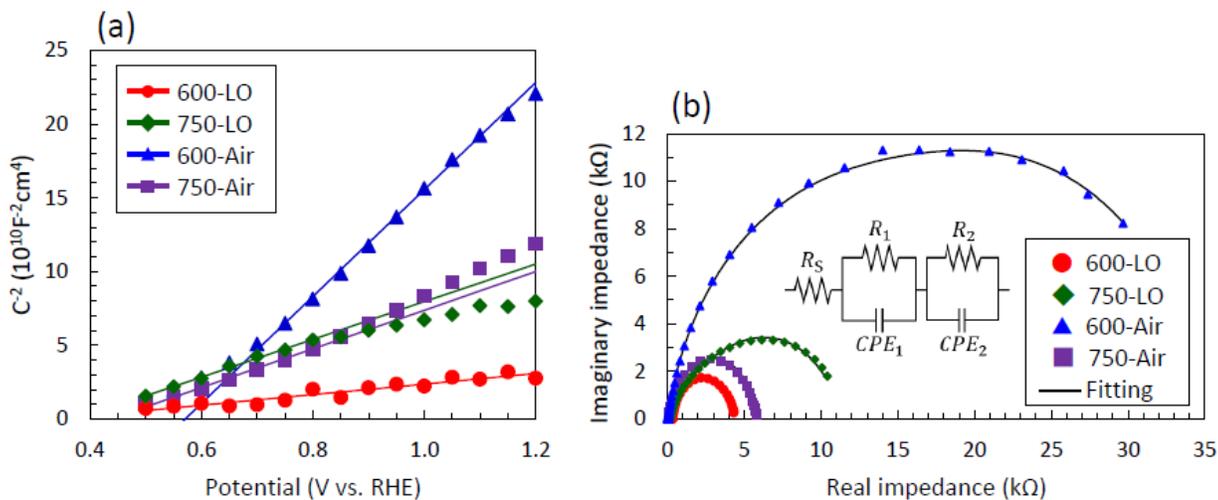

Figure 5. (a) Mott-Schottky plots measured under dark condition and (b) EIS Nyquist plots measured under illumination at 1.5 V vs. RHE for α-Fe$_2$O$_3$ layers annealed under various conditions. The inset is an equivalent circuit model for a fitting the experimental data.

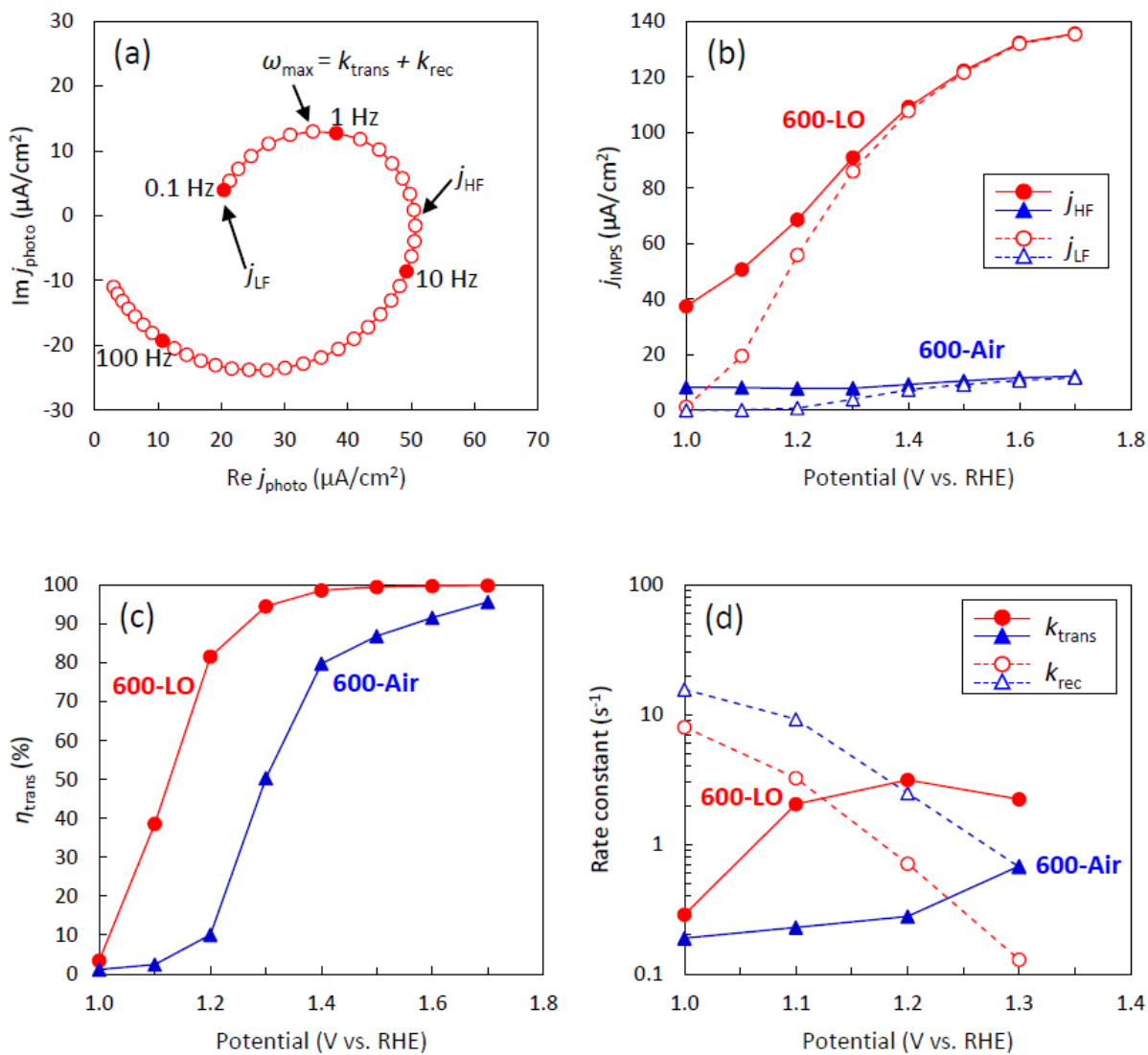

Figure 6. (a) Typical IMPS response for 600-LO at 1.1 V vs. RHE, variations of (b) low frequency (LF) and high frequency (HF) limits of IMPS response, (c) transfer efficiency, and (d) rate constants of electron-hole recombination and hole transfer as a function of applied potential.

Table 1. Sn atomic ratio evaluated from Sn 3d XPS data, $Fe^{2+}/Fe^{3+}$ ratio evaluated from Fe $2p_{3/2}$ satellite peak area, resistance collected by conductivity measurements, and donor

density evaluated from slopes of Mott-Schottky plots for α-Fe$_2$O$_3$ layers annealed under various conditions.

|  | 750-LO | 600-LO | 750-Air | 600-Air |
|---|---|---|---|---|
| Sn atomic ratio (%) | 6.0 | 3.8 | 2.1 | 1.1 |
| Fe$^{2+}$/Fe$^{3+}$ | 1.32 | 1.06 | 0.83 | 0.77 |
| Resistance (MΩ) | 2.64 | 0.33 | 9.84 | 39.91 |
| $N_d$ (cm$^{-3}$) | 1.38 × 10$^{+19}$ | 4.88 × 10$^{+19}$ | 1.34 × 10$^{+19}$ | 4.86 × 10$^{+18}$ |